\documentclass{elsart}
\usepackage{ifpdf}
\usepackage{graphicx,amssymb,lineno,epsfig,subfig}
\usepackage{subfig}

\usepackage{indentfirst}

\ifpdf
\usepackage[%
  pdftitle={Instructions for use of the document class
    elsart},%
  pdfauthor={Simon Pepping},%
  pdfsubject={The preprint document class elsart},%
  pdfkeywords={instructions for use, elsart, document class},%
  pdfstartview=FitH,%
  bookmarks=true,%
  bookmarksopen=true,%
  breaklinks=true,%
  colorlinks=true,%
  linkcolor=blue,anchorcolor=blue,%
  citecolor=blue,filecolor=blue,%
  menucolor=blue,pagecolor=blue,%
  urlcolor=blue]{hyperref}
\else
\usepackage[%
  breaklinks=true,%
  colorlinks=true,%
  linkcolor=blue,anchorcolor=blue,%
  citecolor=blue,filecolor=blue,%
  menucolor=blue,pagecolor=blue,%
  urlcolor=blue]{hyperref}
\fi

\renewcommand{\theequation}{\arabic{section}.\arabic{equation}}
\newtheorem{theorem}{Theorem}[section]
\newtheorem{remark}{Remark}[section]

\newtheorem{lemma}{Lemma}[section]
\newenvironment{proof}[1][Proof]{\noindent\emph{{#1.}} }{ \rule{0.5em}{0.5em}}
\usepackage{indentfirst}
\usepackage{setspace}

\begin{document}
\begin{frontmatter}
\title{Solitons, peakons and periodic cusp wave solutions for the Fornberg-Whitham equation}
\author{Jiangbo Zhou\corauthref{cor}},
\corauth[cor]{Corresponding author. Tel.: +86-511-88969336; Fax:
+86-511-88969336.} \ead{zhoujiangbo@yahoo.cn}
\author{Lixin Tian}
\address{Nonlinear Scientific Research Center, Faculty of Science, Jiangsu
University, Zhenjiang, Jiangsu 212013, China}
\begin{abstract} In this paper, we employ the bifurcation method of dynamical systems
to investigate the exact travelling wave solutions for the
Fornberg-Whitham equation $u_t - u_{xxt} + u_x + uu_x= uu_{xxx}+
3u_x u_{xx}$. The implicit expression for solitons is given. The
explicit expressions for peakons and periodic cusp wave solutions
are also obtained. Further, we show that the limits of soliton
solutions and periodic cusp wave solutions are peakons.
\end{abstract}


\end{frontmatter}
\section{Introduction}
 \setcounter {equation}{0}
The Fornberg-Whitham equation
\begin{equation}
\label {eq1.1}u_t - u_{xxt} + u_x + uu_x= uu_{xxx}+ 3u_x u_{xx},
\end{equation}
has appeared in the study of qualitative behaviors of wave breaking
\cite {1, 2}. It is a nonlinear dispersive wave equation. Since
Eq.(\ref{eq1.1}) was derived, little attention has been paid to
studying it. In \cite {3}, Fornberg and Whitham obtained a peaked
solution of the form $u(x,t)=A \exp{(-\frac{1}{2}\left| {x -
\frac{4}{3}t} \right|)}$, where $A$ is an arbitrary constant. In
\cite {4}, we constructed a type of bounded travelling wave
solutions for Eq.(\ref{eq1.1}), which are called kink-like and
antikink-like wave solutions. Unfortunately, the results in \cite{3,
4} are not complete. In the present paper, we continue to derive
more travelling wave solutions for Eq.(\ref{eq1.1}), so that we can
supplement the results of \cite{3, 4}.

The remainder of the paper is organized as follows. In Section 2, we
discuss the bifurcation curves and phase portraits of travelling
wave system. In Section 3, we obtain the implicit expression for
solitons and the explicit expressions for peakons and periodic cusp
wave solutions. At the same time, we show that the limits of
solitons and periodic cusp wave solutions are peakons. A short
conclusion is given in Section 4.

\section{Bifurcation and phase portraits of travelling wave
system}
 \setcounter {equation}{0}
Let $u = \varphi (\xi )$ with $\xi = x - ct$ be the solution for
Eq.(\ref{eq1.1}); then it follows that
\begin{equation}
\label{eq2.1}
 - c\varphi' + c\varphi ''' + \varphi '+\varphi
\varphi '  = \varphi \varphi ''' + 3\varphi '\varphi ''.
\end{equation}

Integrating Eq. (\ref{eq2.1}) once we have
\begin{equation}
\label{eq2.2} \varphi ''(\varphi - c) = g - c\varphi + \varphi +
\frac{1}{2}\varphi ^2 - (\varphi ')^2,
\end{equation}
\noindent where $g$ is the integral constant.

Let $y = \varphi '$; then we get the following planar dynamical
system:
\begin{equation}
\label{eq2.3} \left\{ {\begin{array}{l}
 \frac{\textstyle d\varphi }{\textstyle  d\xi } = y ,\\
 \frac{\textstyle dy}{\textstyle  d\xi } = \frac{\textstyle g - c\varphi + \varphi + \frac{1}{2}\varphi ^2 -
y^2}{\textstyle \varphi - c},
 \end{array}} \right.
\end{equation}
\noindent with a first integral
\begin{equation}
\label{eq2.4} H(\varphi ,y) = (\varphi - c)^2(y^2 -
\frac{1}{4}\varphi^2 + (\frac{1}{2}c-\frac{2}{3})\varphi
+\frac{1}{4} c^2-  \frac{1}{3}c- g) = h,
\end{equation}
\noindent where $h$ is a constant.

Note that (\ref{eq2.3}) has a singular line $\varphi = c$. To avoid
the line temporarily we make transformation $d\xi = (\varphi -
c)d\zeta $. Under this transformation, Eq.(\ref{eq2.3}) becomes
\begin{equation}
\label{eq2.5} \left\{ {\begin{array}{l}
 \frac{\textstyle d\varphi }{\textstyle d\zeta } = (\varphi - c)y ,\\
 \frac{\textstyle dy}{\textstyle d\zeta } = g - c\varphi + \varphi + \frac{1}{2}\varphi ^2 -
 y^2.
\\
 \end{array}} \right.
\end{equation}

System (\ref{eq2.3}) and system (\ref{eq2.5}) have the same first
integral as (\ref{eq2.4}). Consequently, system (\ref{eq2.5}) has
the same topological phase portraits as system (\ref{eq2.3}) except
for the straight line $\varphi = c$. Obviously, $\varphi = c$ is an
invariant straight-line solution for system (\ref{eq2.5}).

 For a fixed $h$, (\ref{eq2.4}) determines a
set of invariant curves of system (\ref{eq2.5}). As $h$ is varied,
(\ref{eq2.4}) determines different families of orbits of system
(\ref{eq2.5}) having different dynamical behaviors. Let $M(\varphi
_e ,y_e )$ be the coefficient matrix of the linearized version of
(\ref{eq2.5}) at the equilibrium point $(\varphi _e ,y_e )$; then
\begin{equation}
\label{eq2.6} M(\varphi _e ,y_e ) = \left( {{\begin{array}{*{20}c}
{\quad\quad y_e } \hfill &&& {\varphi _e - c} \hfill \\
 {\varphi _e - (c - 1)} \hfill &&&{- 2y_e} \hfill \\
\end{array} }} \right)
\end{equation}
and at this equilibrium point, we have
\begin{equation}
\label{eq2.7} J(\varphi _e ,y_e ) = \det M(\varphi _e ,y_e ) = -
2y_e^2 - (\varphi _e - c)[\varphi _e - (c - 1)],
\end{equation}
\begin{equation}
\label{eq2.8} p(\varphi _e ,y_e ) = \mathrm{trace}(M(\varphi _e ,y_e
)) = - y_e.
\end{equation}
By the theory of planar dynamical systems (see \cite {5}), for an
equilibrium point of a planar dynamical system, if $J < 0$, then
this equilibrium point is a saddle point; it is a center point if $J
> 0$ and $p = 0$; if $J = 0$ and the Poincar\'{e} index of the
equilibrium point is 0, then it is a cusp.

By using the first-integral value and properties of equilibrium
points, we obtain the bifurcation curves as follows:
\begin{equation}
\label{eq2.9} g_1(c)=\frac{1}{2}(c-1)^2,
\end{equation}

\begin{equation}
\label{eq2.10} g_2(c)=\frac{1}{2}(c-1)^2-\frac{1}{18},
\end{equation}

\begin{equation}
\label{eq2.11} g_3(c)=\frac{1}{2}(c-1)^2-\frac{1}{2}.
\end{equation}
Obviously, the three curves have no intersection point and
$g_3(c)<g_2(c)<g_1(c)$ for arbitrary constant $c$.

Using the bifurcation method for vector fields (e.g., \cite{5}), we
have the following result which describes the locations and
properties of the singular points of system (\ref{eq2.5}).
\begin{theorem}
For given any constant wave speed $c\neq 0$, let
\begin{equation}
\label{eq2.12} \varphi _{1\pm} = c - 1\pm \sqrt {(c - 1)^2 -
2g}\quad for \quad g\leq g_1(c),
\end{equation}
\begin{equation}
\label{eq2.13} y _{1\pm} =\pm \sqrt {g-\frac{1}{2}c^2+c} \quad for
\quad g\geq g_3(c).
\end{equation}
Then we have

(1)If $g<g_3(c)$, then system (\ref{eq2.5}) has two equilibrium
points $(\varphi _{1 -} ,0)$ and $(\varphi _{1+} ,0)$, which are
saddle points.

(2)If $g=g_3(c)$, then system (\ref{eq2.5}) has two equilibrium
points $(c-2, 0)$ and $(c,0)$. $(c-2, 0)$ is a saddle point and
$(c,0)$ is a cusp.

(3)If $g_3(c)<g<g_2(c)$, then system (\ref{eq2.5}) has four
equilibrium points $(\varphi _{1-} ,0)$, $(\varphi _{1+} ,0)$,
$(c,y_{1-})$ and
 $(c,y_{1+})$. $(\varphi _{1-},0)$ is a saddle point and $(\varphi _{1+} ,0)$
 is a center point enclosing the orbit which connects the saddle
 points $(c,y_{1-})$ and  $(c,y_{1+})$.

(4)If $g=g_2(c)$, then system (\ref{eq2.5}) has four  equilibrium
points $(c-\frac{4}{3}, 0)$, $(c-\frac{2}{3},0)$, $(c,-\frac{2}{3})$
and $(c,\frac{2}{3})$, which satisfy $H(c-\frac{4}{3},
0)=H(c,-\frac{2}{3})=H(c,\frac{2}{3})$ and form a triangular orbit
which encloses the center point $(c-\frac{2}{3},0)$.

(5)If $g_2(c)<g<g_1(c)$, then system (\ref{eq2.5}) has four
equilibrium points $(\varphi _{1-} ,0)$, $(\varphi _{1+} ,0)$,
$(c,y_{1-})$ and
 $(c,y_{1+})$. $(\varphi _{1+} ,0)$ is a center point enclosing the orbit which is homoclinic for the saddle point
 $(\varphi _{1-}, 0)$.

(6)If $g=g_1(c)$, then system (\ref{eq2.5}) has three  equilibrium
points $(c-1, 0)$, $(c,-\frac{\sqrt{2}}{2})$ and
$(c,\frac{\sqrt{2}}{2})$. $(c-1, 0)$ is a cusp.
$(c,-\frac{\sqrt{2}}{2})$ and $(c,\frac{\sqrt{2}}{2})$ are two
saddle points.

(7)If $g>g_1(c)$, then system (\ref{eq2.5}) has two equilibrium
points $(c,y_{1-})$ and $(c,y_{1+})$. They are saddle points.

\end{theorem}

\begin{figure}[h]
\centering
\subfloat[]
{\includegraphics[height=1.0in,width=1.2in]{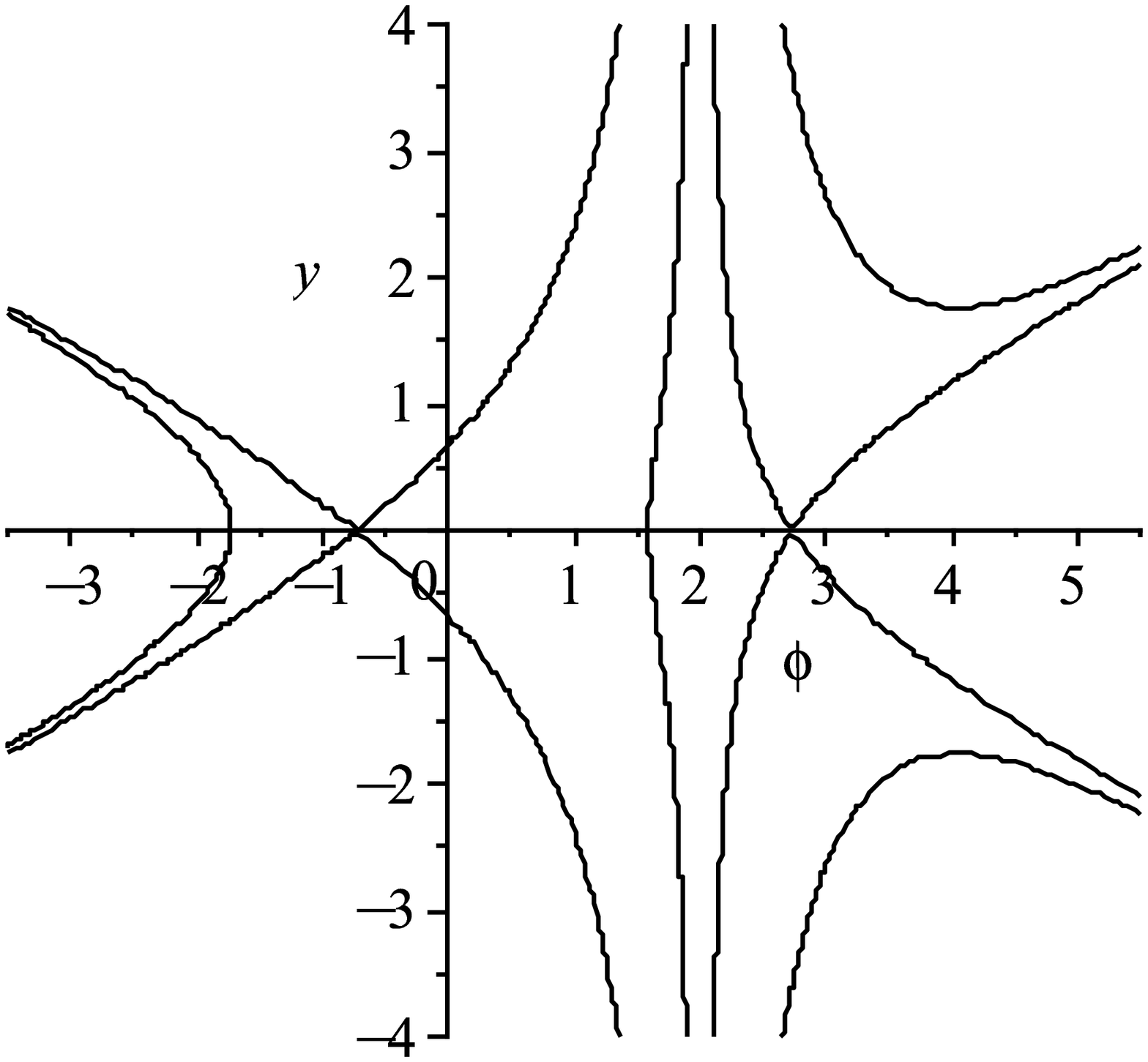}}\hspace{0.02\linewidth}
\subfloat[]{\includegraphics[height=1.0in,width=1.2in]{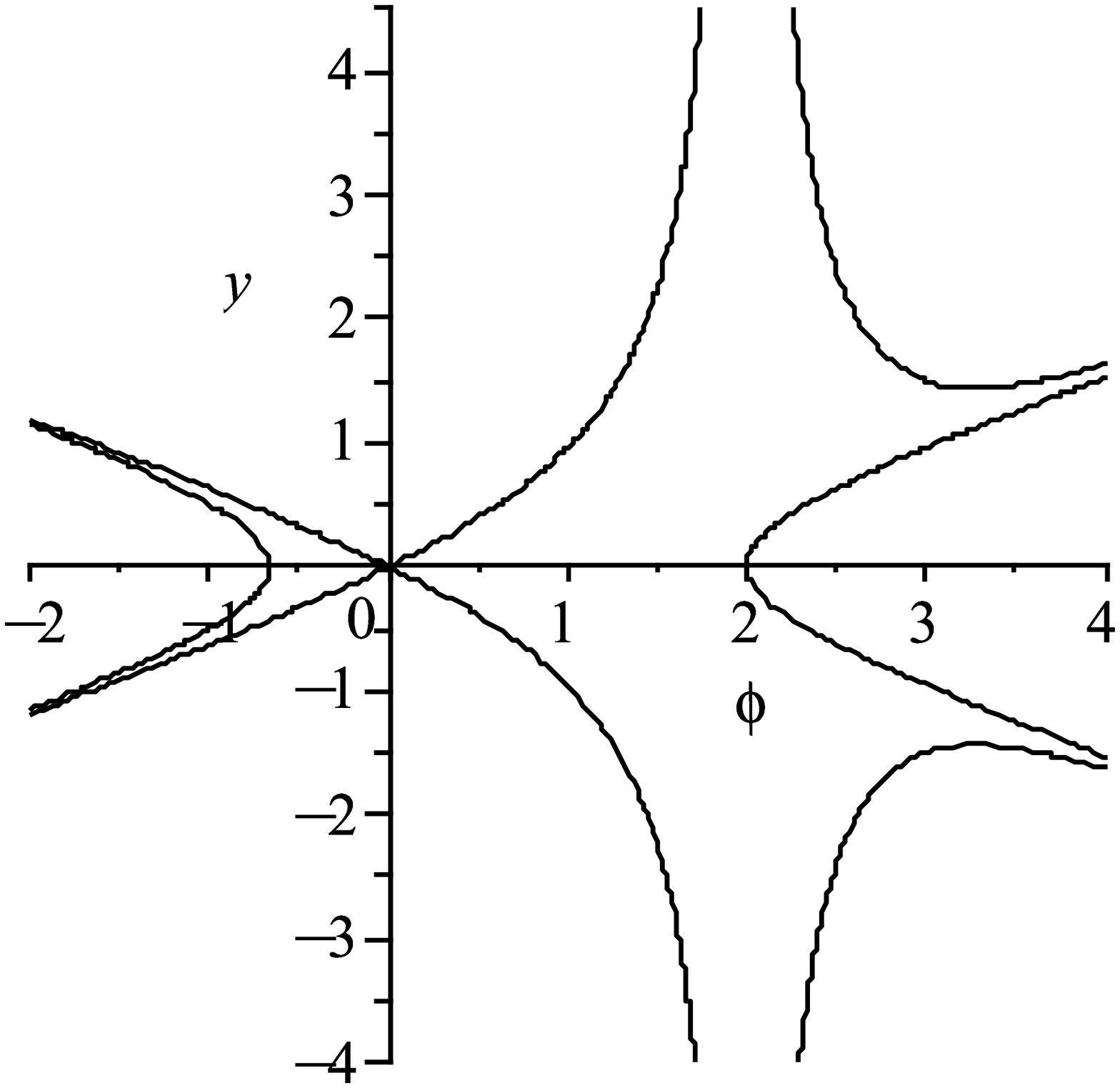}}\hspace{0.02\linewidth}
\subfloat[]{\includegraphics[height=1.0in,width=1.2in]{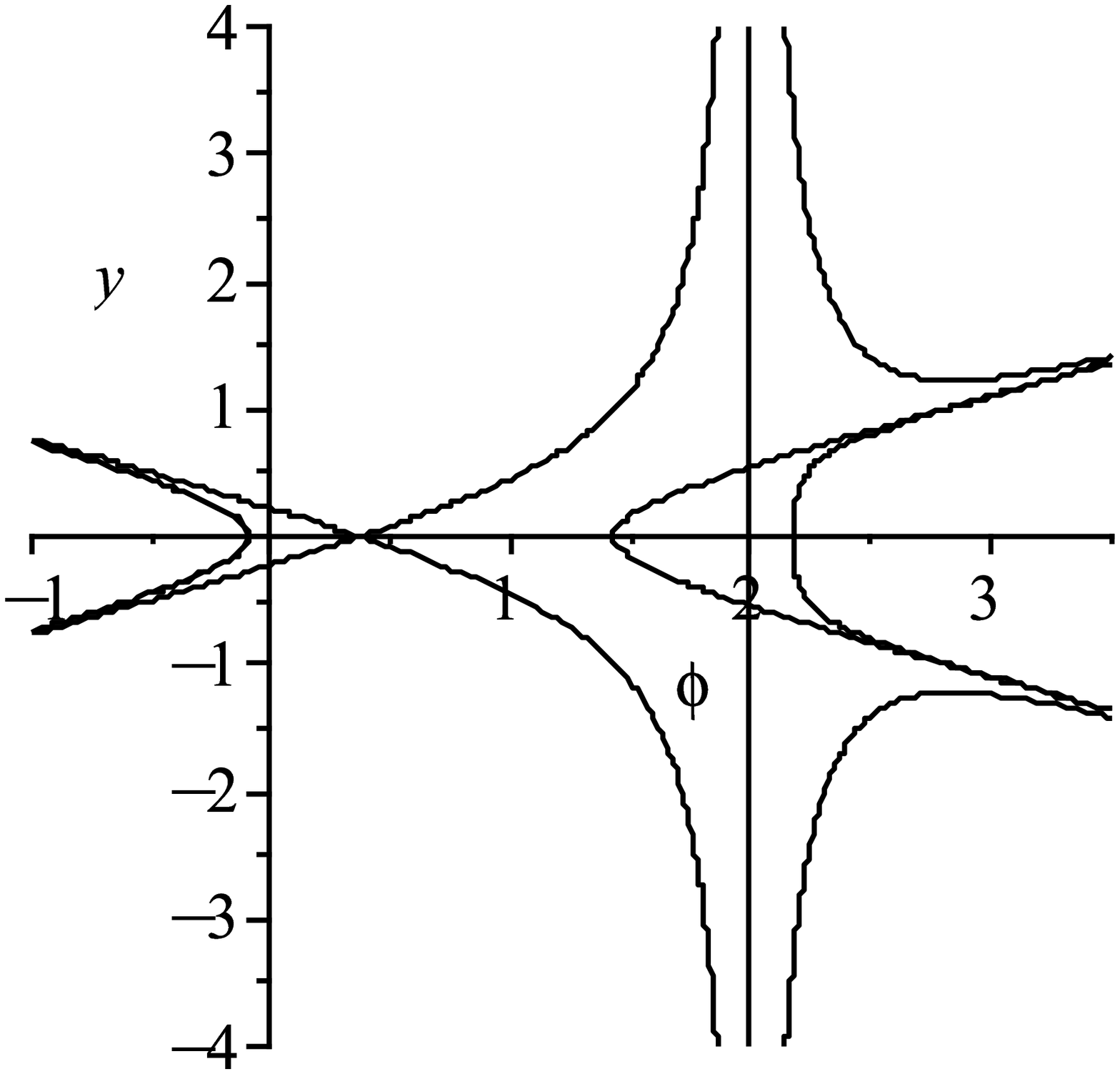}}\hspace{0.02\linewidth}
\subfloat[]{\includegraphics[height=1.0in,width=1.2in]{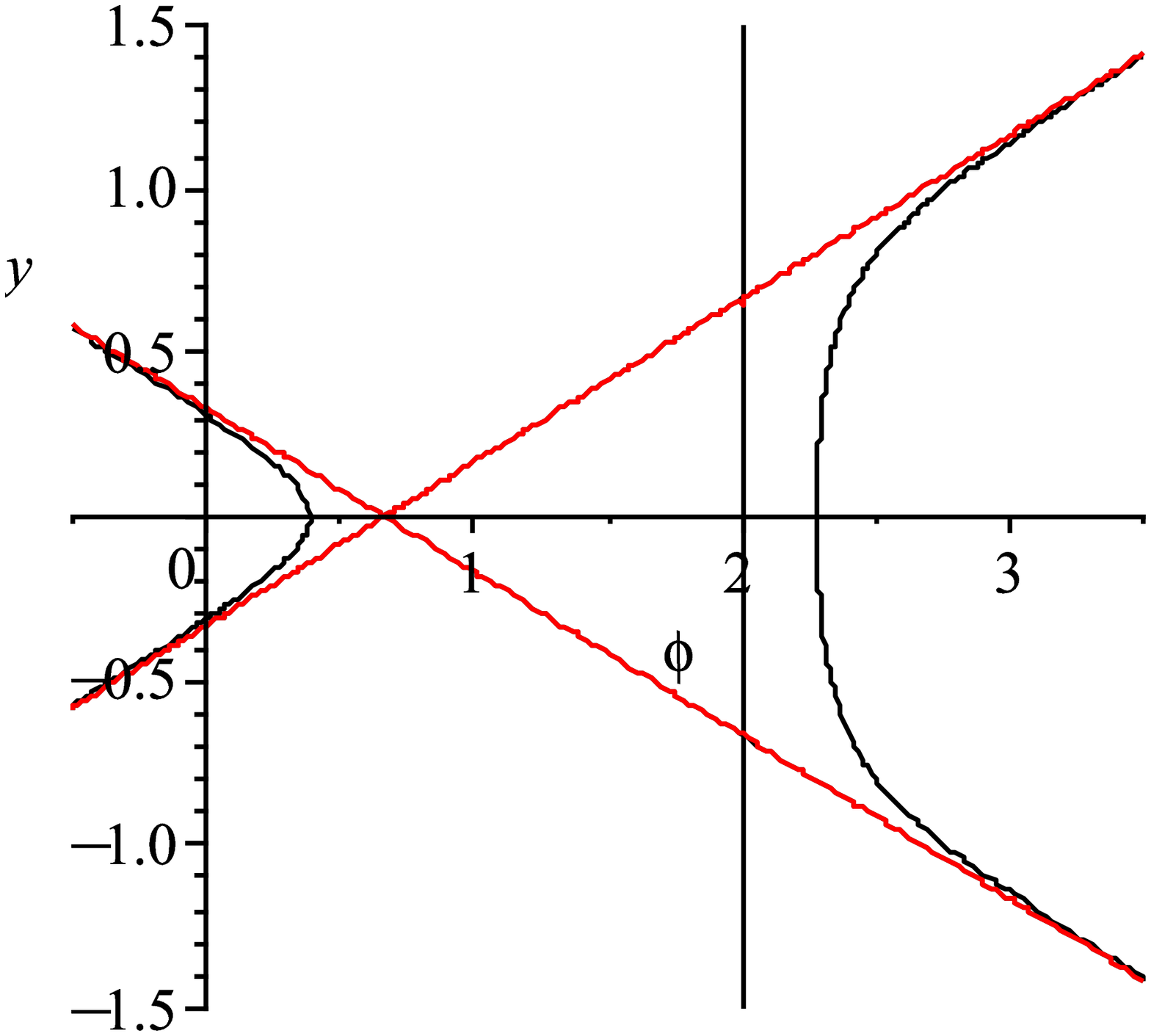}}\\
\subfloat[]{\includegraphics[height=1.0in,width=1.2in]{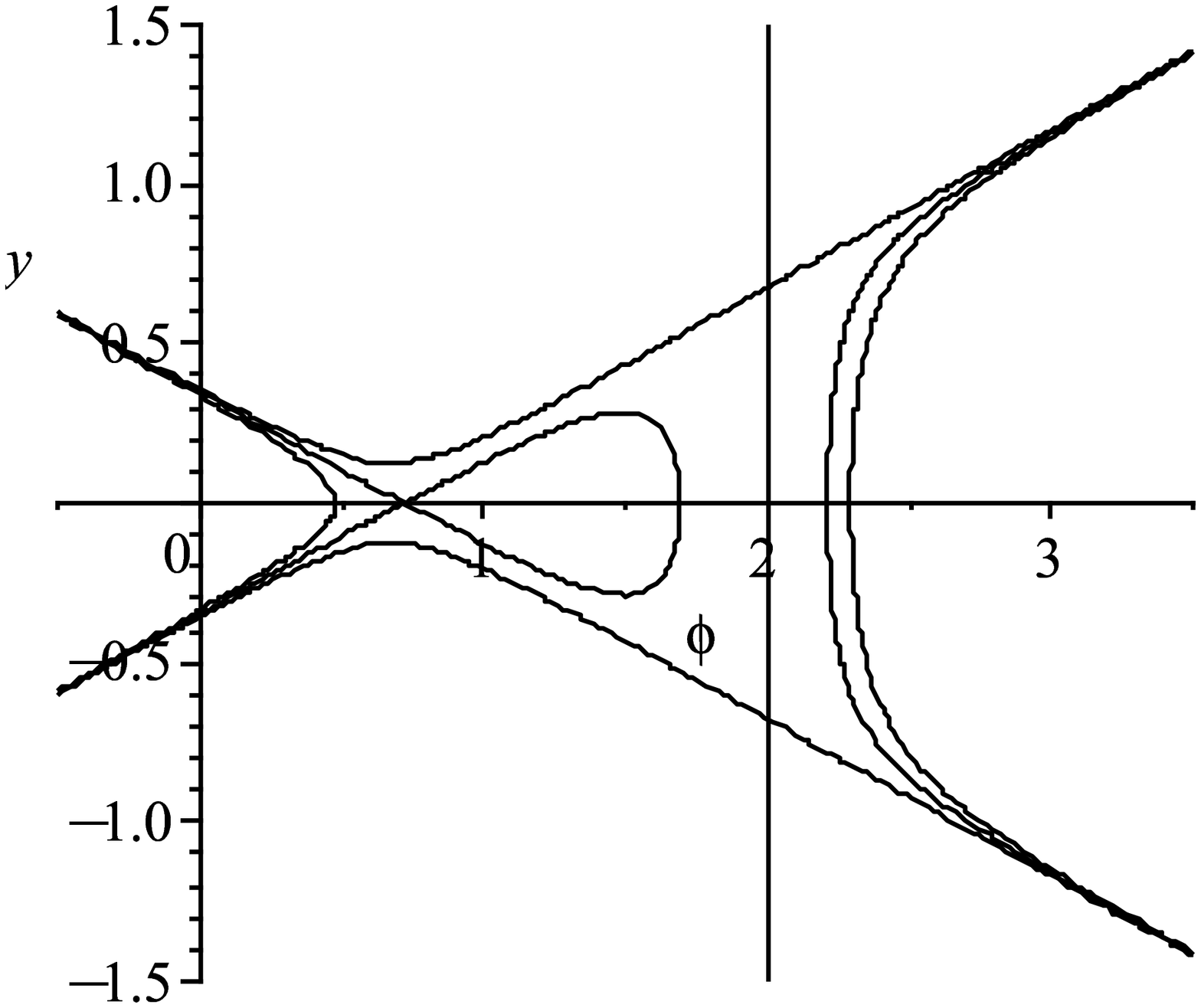}}\hspace{0.07\linewidth}
\subfloat[]{\includegraphics[height=1.0in,width=1.2in]{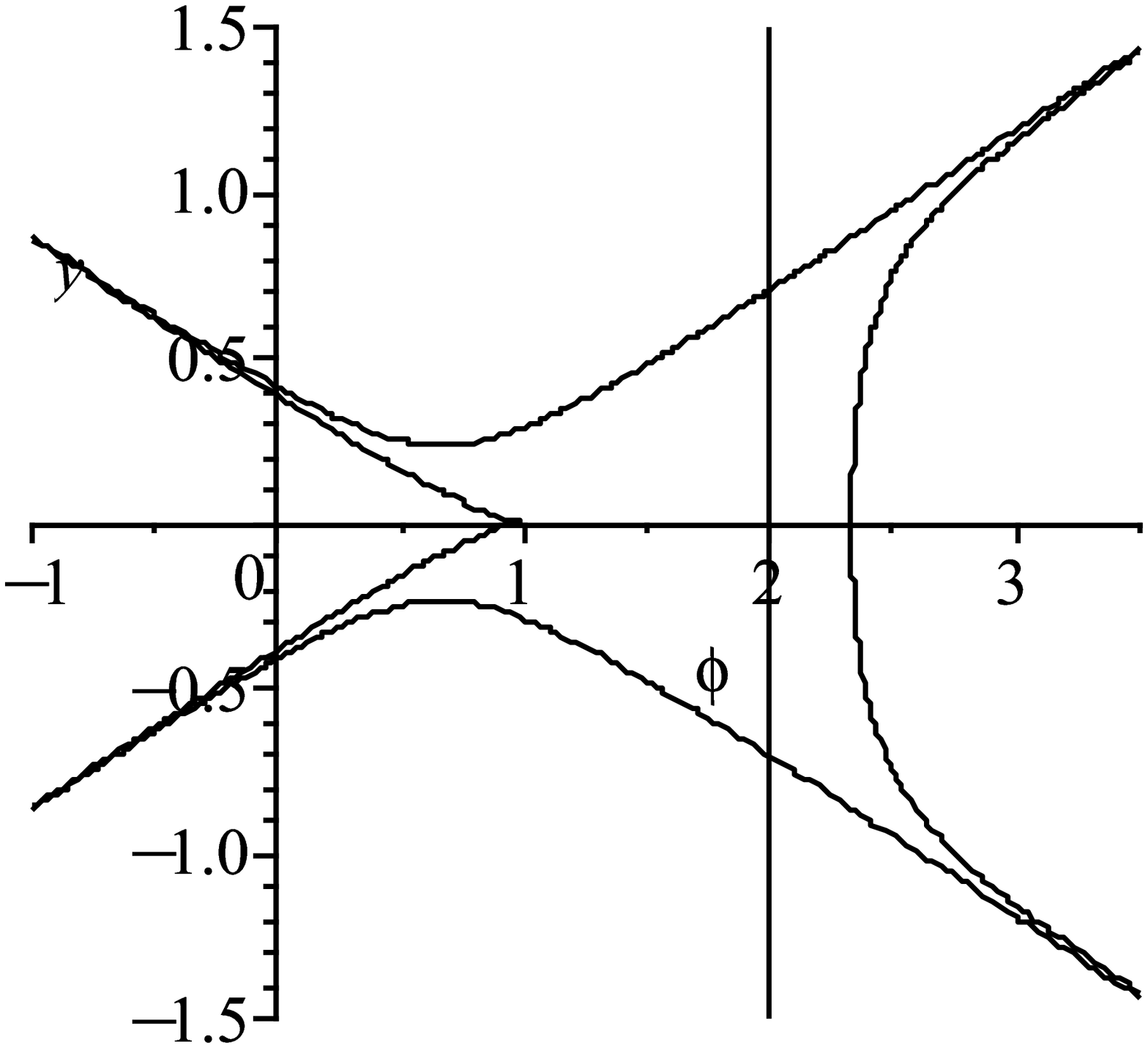}}\hspace{0.07\linewidth}
\subfloat[]{\includegraphics[height=1.0in,width=1.2in]{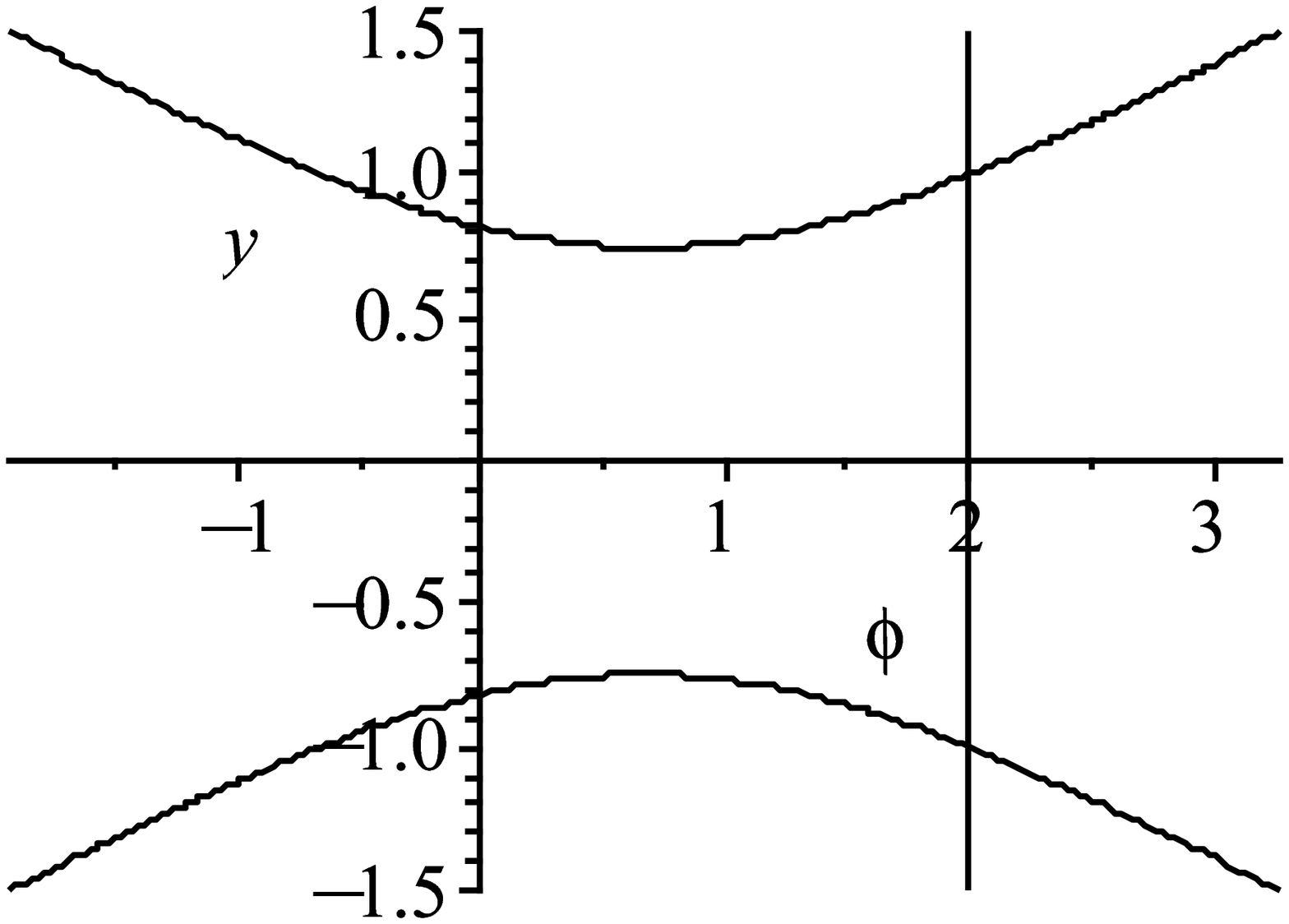}}

\caption{The phase portraits of system (\ref{eq2.5}). (a)
$g<g_3(c)$; (b) $g=g_3(c)$; (c) $g_3(c)<g<g_2(c)$; (d) $g=g_2(c)$;
(e) $g_2(c)<g<g_1(c)$; (f) $g=g_1(c)$; (g) $g>g_1(c)$. }\label{f1}
\end{figure}

The phase portraits of system (\ref{eq2.5}) are given in
Fig.\ref{f1}.

\section{Solitons, peakons and periodic cusp wave solutions}
\begin{figure}[h]
\centering
\subfloat[]{\includegraphics[height=1.3in,width=1.4in]{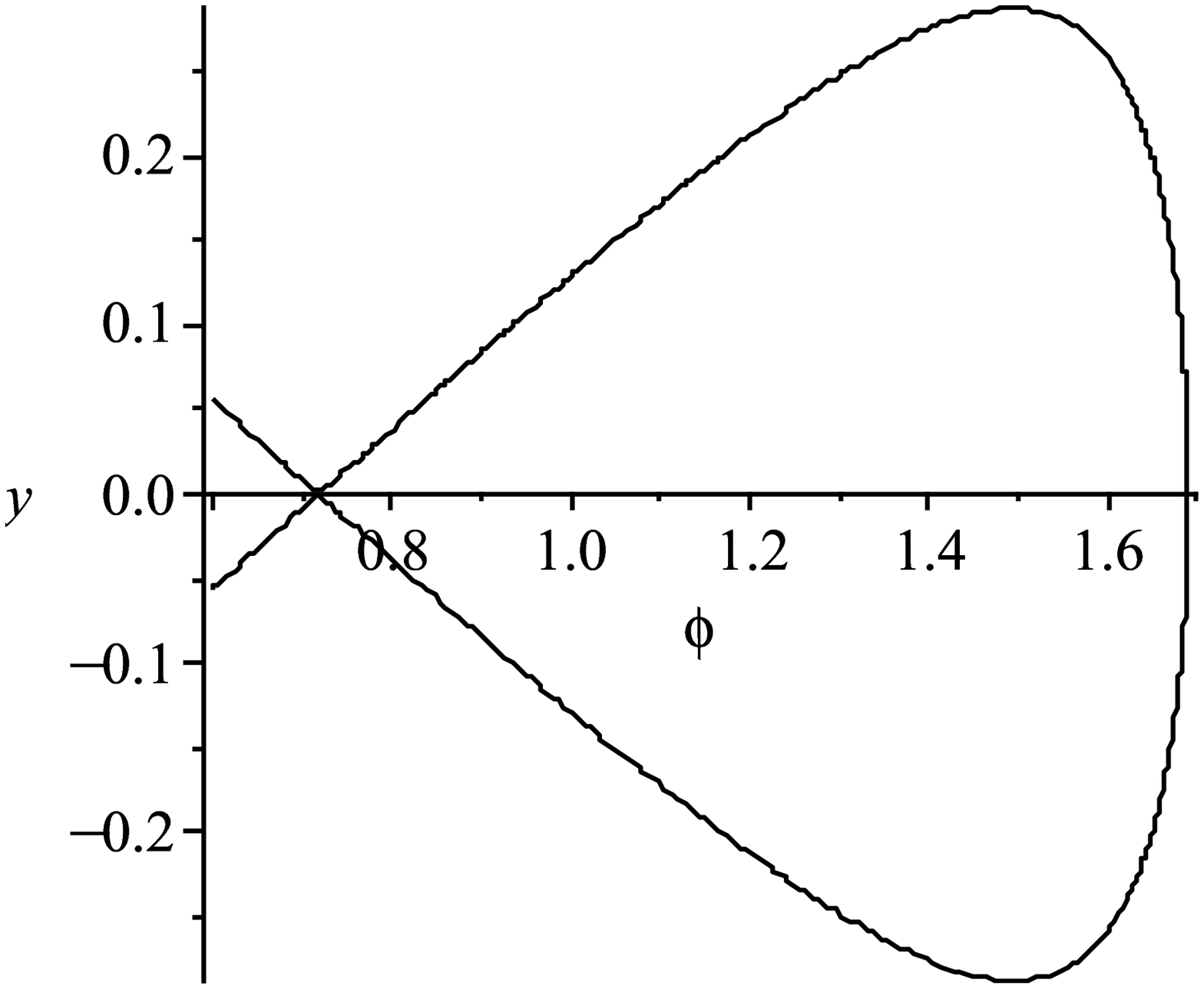}}\hspace{0.04\linewidth}
\subfloat[]{\includegraphics[height=1.2in,width=1.6in]{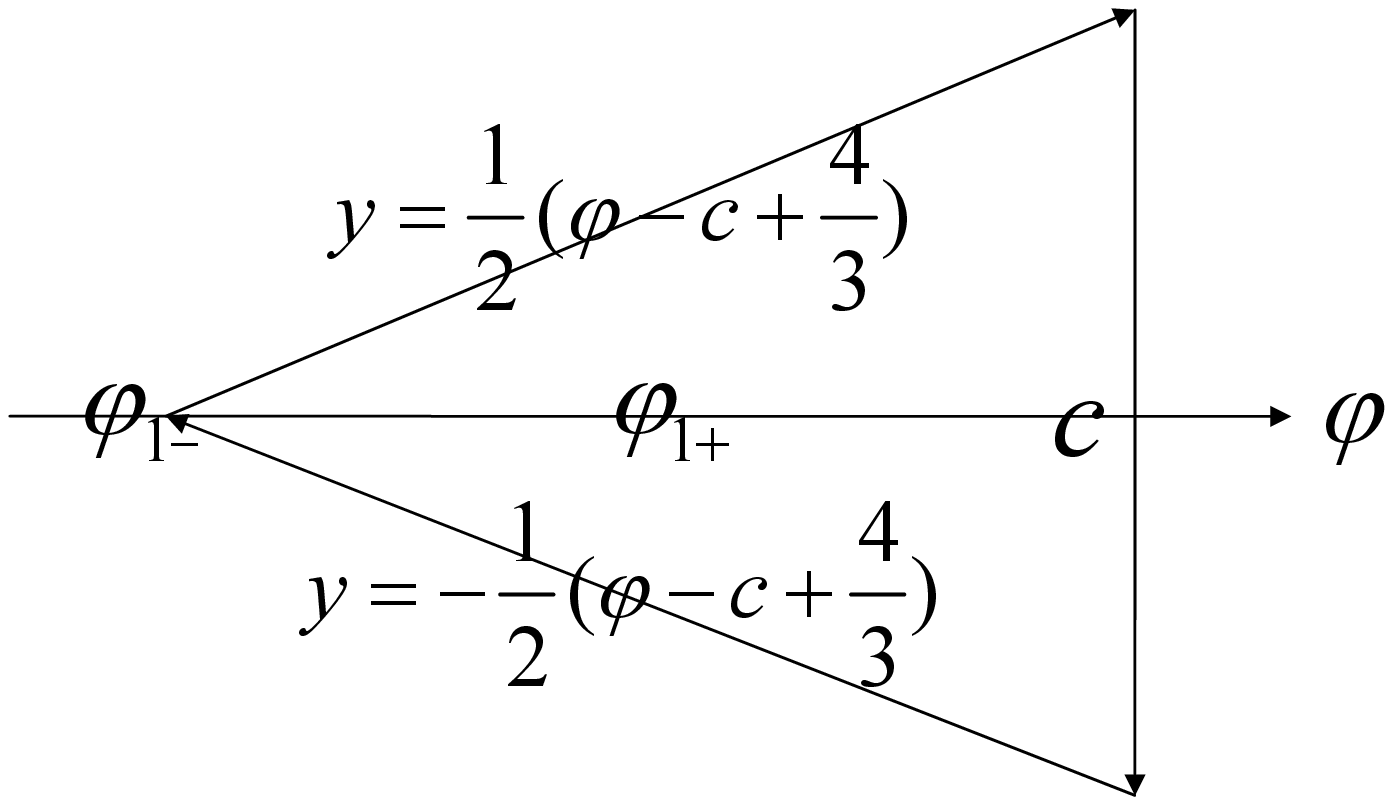}}\hspace{0.04\linewidth}
\subfloat[]{\includegraphics[height=1.2in,width=1.6in]{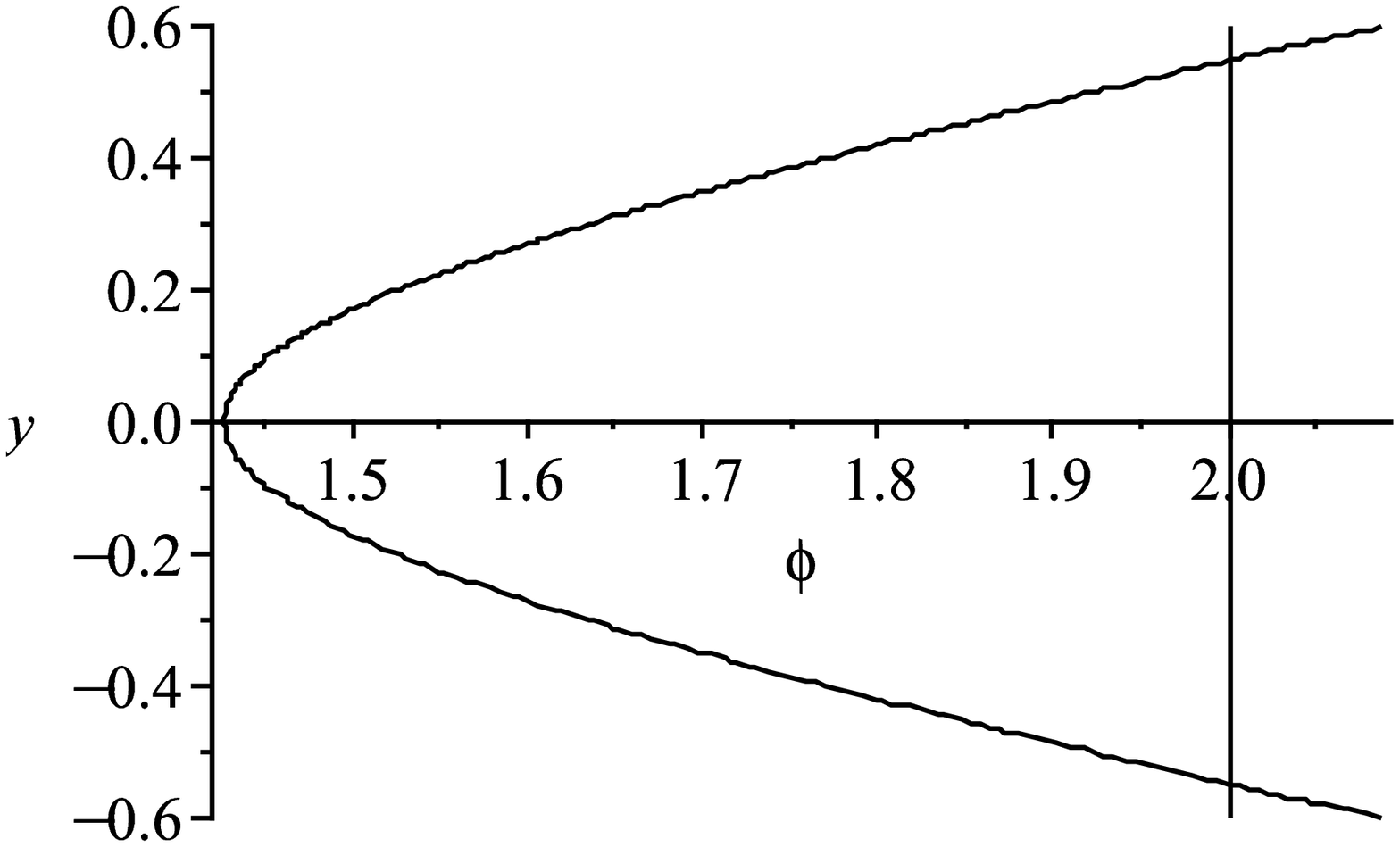}}
\caption{The orbits of system (\ref{eq2.3}). (a) The homoclinic
orbit (corresponding to $g_2(c)<g<g_1(c)$). (b) The limit curve of
homoclinic orbit and periodic orbit (corresponding to $g=g_2(c)$).
(c) The periodic orbit (corresponding to
$g_3(c)<g<g_2(c)$).}\label{f2}
\end{figure}

 \setcounter {equation}{0}
    Suppose that $\varphi(\xi )(\xi = x - ct)$ is a travelling wave
solution for Eq.(\ref{eq1.1}) for $\xi\in ( - \infty, + \infty )$,
and $\mathop {\lim }\limits_{\xi \to - \infty } \varphi(\xi ) = A$,
$\mathop {\lim }\limits_{\xi\to\infty } \varphi(\xi ) = B$, where
$A$ and $B$ are two constants. If $A=B$, then $\varphi(\xi )$ is
called a soliton solution. If $A \ne B$, then $\varphi(\xi )$ is
called a kink (or an antikink) solution. Usually, a soliton solution
for Eq.(\ref{eq1.1}) corresponds to a homoclinic orbit of system
(\ref{eq2.3}) and a periodic travelling wave solution for
Eq.(\ref{eq1.1}) corresponds to a periodic orbit of system
(\ref{eq2.3}). Similarly, a kink (or an antikink) wave solution of
Eq.(\ref{eq1.1}) corresponds to a heteroclinic orbit (or the
so-called connecting orbit) of  system (\ref{eq2.3}). The graphs of
the  homoclinic orbit, periodic orbit and their limit cure are shown
in Fig.\ref{f2}.

The following lemma gives the relationship of soliton solutions of
Eq.(\ref{eq1.1}) and homoclinic orbits of system (\ref{eq2.3}).
\begin{lemma}
\label{le3.1} Assume that $\Gamma$ is a homoclinic orbit of system
(\ref{eq2.3}) and its parameter expression is $\varphi=\varphi(\xi)$
and $y=y(\xi)$; then $u=\varphi(\xi)$ with $\xi=x-ct$ is a soliton
solution for Eq.(\ref{eq1.1}).
\end{lemma}
\begin{proof}  From Fig.\ref{f1}(e), we can see that the homoclinic orbit
$\Gamma$ encloses $(\varphi_{1+},0)$ and connects
$(\varphi_{1-},0)$. Therefore,
$\lim_{|\xi|\rightarrow\infty}\varphi(\xi)=\varphi_{1-}$.

On the other hand, $u=\varphi(\xi)$ is the solution for system
(\ref{eq2.3}). This implies that  $u=\varphi(\xi)$ is the solution
for Eq.(\ref{eq2.1}). Thus, $u=\varphi(x-ct)$ is the soliton
solution for Eq.(\ref{eq1.1}).
\end{proof}

In Fig.\ref{f2}(a), the homoclinic orbit of system (\ref{eq2.3}) can
be expressed as
\begin{equation}
\label{eq3.1}  y = \pm\frac{( \varphi-\varphi _{1-})\sqrt {\varphi
^2 + l_1 \varphi + l_2 } }{2(\varphi-c)} \quad for \quad \varphi
_{1-}<\varphi <\varphi _{2+},
\end{equation}
where
\begin{equation}
\label{eq3.2} l_1 = \frac{2}{3}(1 - 3c - 3\sqrt {(c - 1)^2 - 2g} ),
\end{equation}
\begin{equation}
\label{eq3.3}l_2 = \frac{2}{3}(1 - 4c + 3c^2 - 3g + (3c + 1)\sqrt
{(c - 1)^2 - 2g} ),
\end{equation}
\begin{equation}
\label{eq3.4}\varphi_{2+}=-\frac{1}{3}(1-3c-3\sqrt {(c - 1)^2 - 2g}
+2\sqrt {1-3\sqrt {(c - 1)^2 - 2g} }).
\end{equation}
Substituting Eq.(\ref{eq3.1}) into the first equation of system
(\ref{eq2.3}) and integrating along the homoclinic orbits, we have
\begin{equation}
\label{eq3.5} {\int_\varphi^{\varphi _{2+}} {\frac{s-c}{(s - \varphi
_{1-} )\sqrt {s^2 + l_1 s + l_2 } }ds = -\frac{1}{2}|\xi| }}.
\end{equation}
It follows from (\ref{eq3.5}) that
\begin{equation}
\label{eq3.6} \beta(\varphi _{2+}) =
\beta(\varphi)\exp(-\frac{1}{2}|\xi|),
\end{equation}
where
\begin{equation}
\label{eq3.7}
 \beta _(\varphi) =  \frac{(2\sqrt {\varphi^2 + l_1
\varphi + l_2 } + 2\varphi + l_1 )(\varphi - \varphi _{1 -
})^{\alpha _1 }}{(2\sqrt {a_1 } \sqrt {\varphi^2 + l_1 \varphi + l_2
} + b_1 \varphi + l_3 )^{\alpha _1 }},
\end{equation}
\begin{equation}
\label{eq3.8} l_1 = \frac{2}{3}(1 - 3c - 3\sqrt {(c - 1)^2 - 2g} ),
\end{equation}
\begin{equation}
\label{eq3.9} l_2 = \frac{2}{3}(1 - 4c + 3c^2 - 3g + (3c + 1)\sqrt
{(c - 1)^2 - 2g}),
\end{equation}
\begin{equation}
\label{eq3.10} l_3 = \frac{4}{3}(2 - 5c + 3c^2 - 6g + (3c + 2)\sqrt
{(c - 1)^2 - 2g} ),
\end{equation}
\begin{equation}
\label{eq3.11} a_1 = 4(1 - 2c + c^2 - 2g + \sqrt {(c - 1)^2 - 2g} ),
\end{equation}
\begin{equation}
\label{eq3.12} b_1 = - \frac{4}{3} - 4\sqrt {(c - 1)^2 - 2g},
\end{equation}
\begin{equation}
\label{eq3.13} \alpha _1 = - \frac{1 + \sqrt {(c - 1)^2 - 2g},
}{2\sqrt {(c - 1)^2 - 2g + \sqrt {(c - 1)^2 - 2g} } }.
\end{equation}

\begin{figure}[h]
\centering
\subfloat[]{\includegraphics[height=1.3in,width=2in]{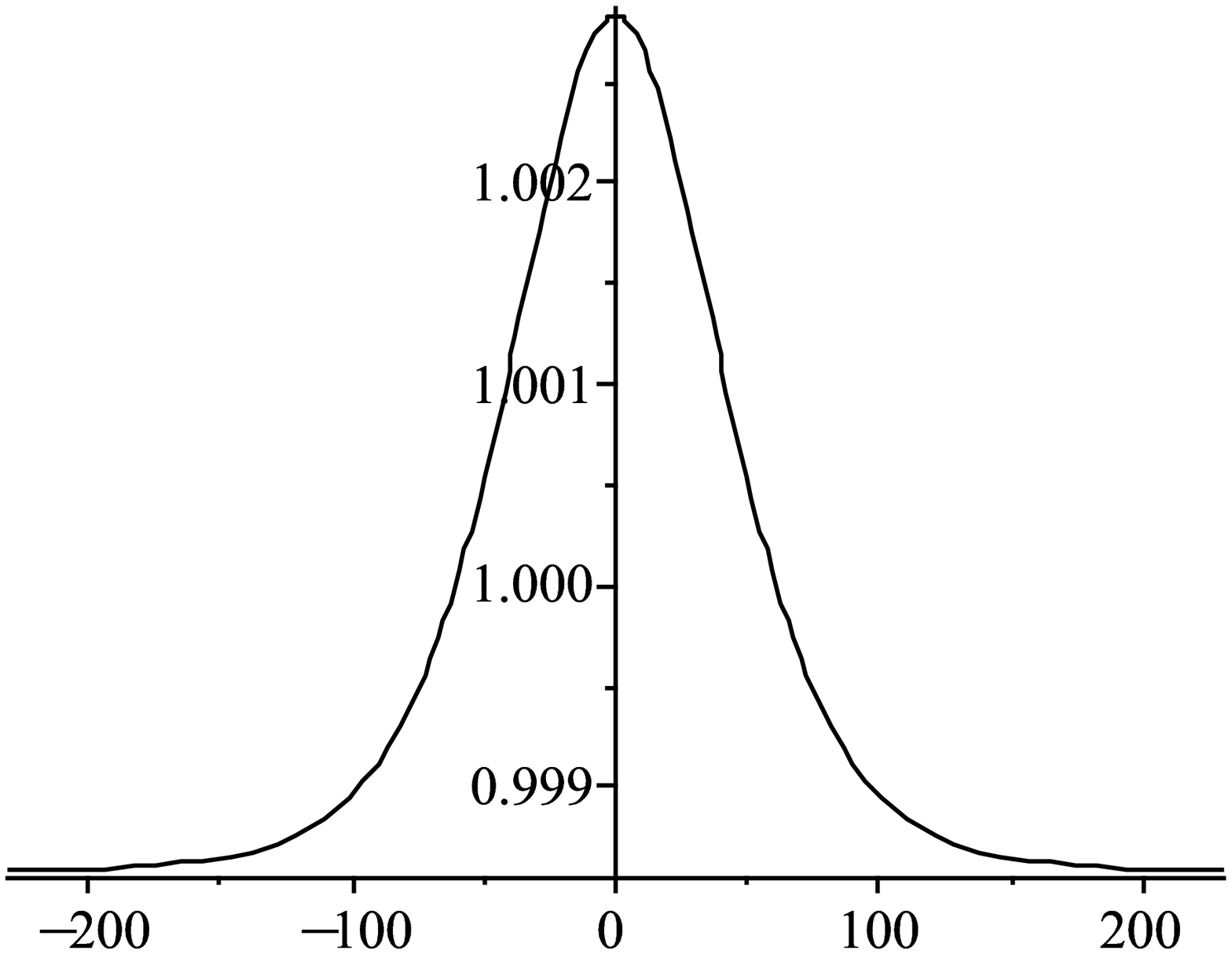}}\hspace{0.15\linewidth}
\subfloat[]{\includegraphics[height=1.3in,width=2in]{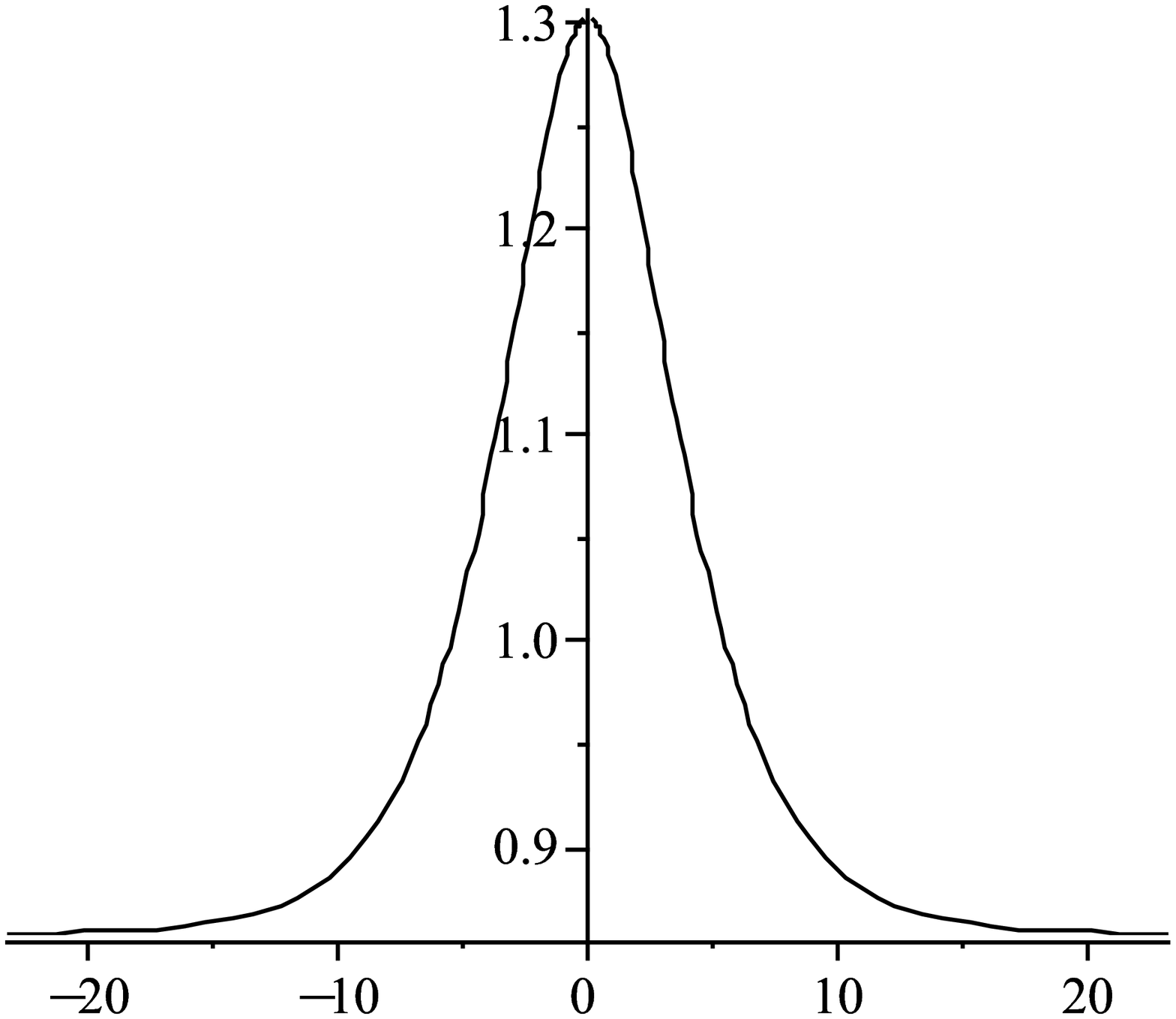}}\\
\subfloat[]{\includegraphics[height=1.3in,width=2in]{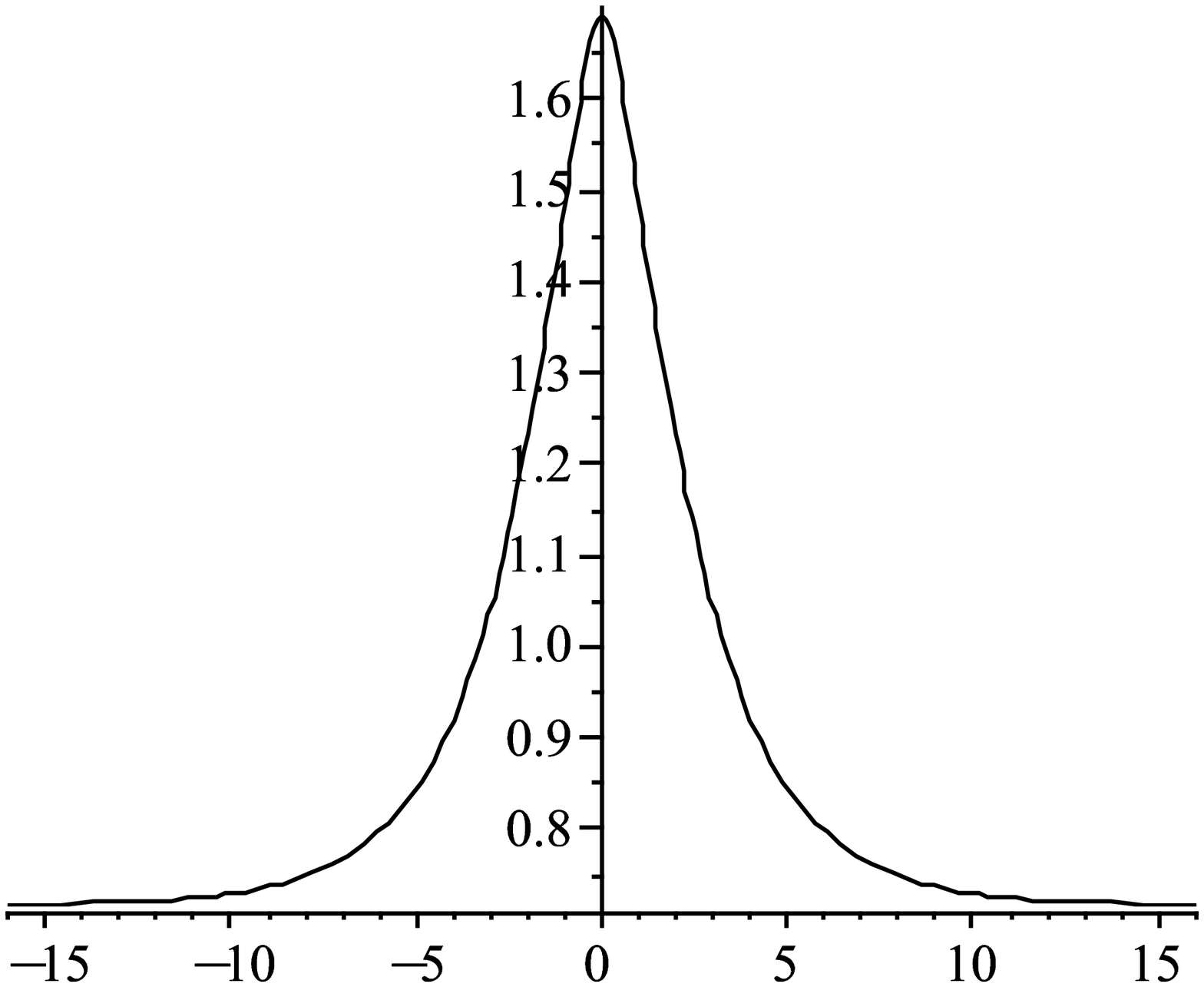}}\hspace{0.15\linewidth}
\subfloat[]{\includegraphics[height=1.3in,width=2in]{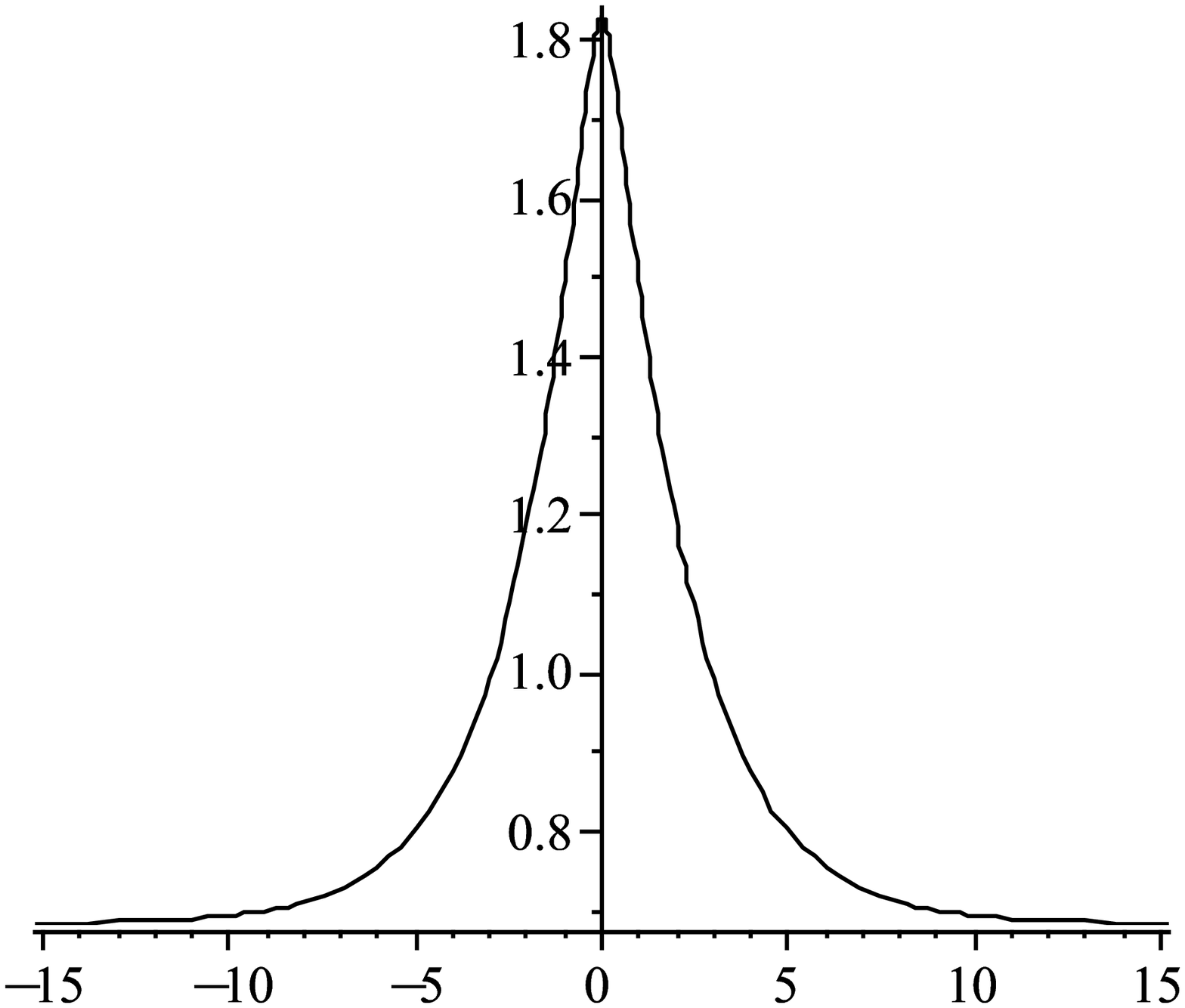}}\\
\caption{The solitons for Eq.(\ref{eq1.1}). (a) $c=2$, $g=0.499999$;
(b) $c=2$, $g=0.49$; (c) $c=2$, $g=0.46$ ; (d) $c=2$,
$g=0.45$.}\label{f3}
\end{figure}

(\ref{eq3.6}) is the implicit expression for solitons for
Eq.(\ref{eq1.1}). We show the graphs of the solitons in Fig.\ref{f3}
under some parameter conditions. From  Fig.\ref{f3}, we can see that
when $g_2(c)<g<g_1(c)$ and $g$ tends to $g_2(c)$, the solitons lose
their smoothness and tend to peakons.

\begin{figure}[h]
\centering
\subfloat[]{\includegraphics[height=1.5in,width=1.5in]{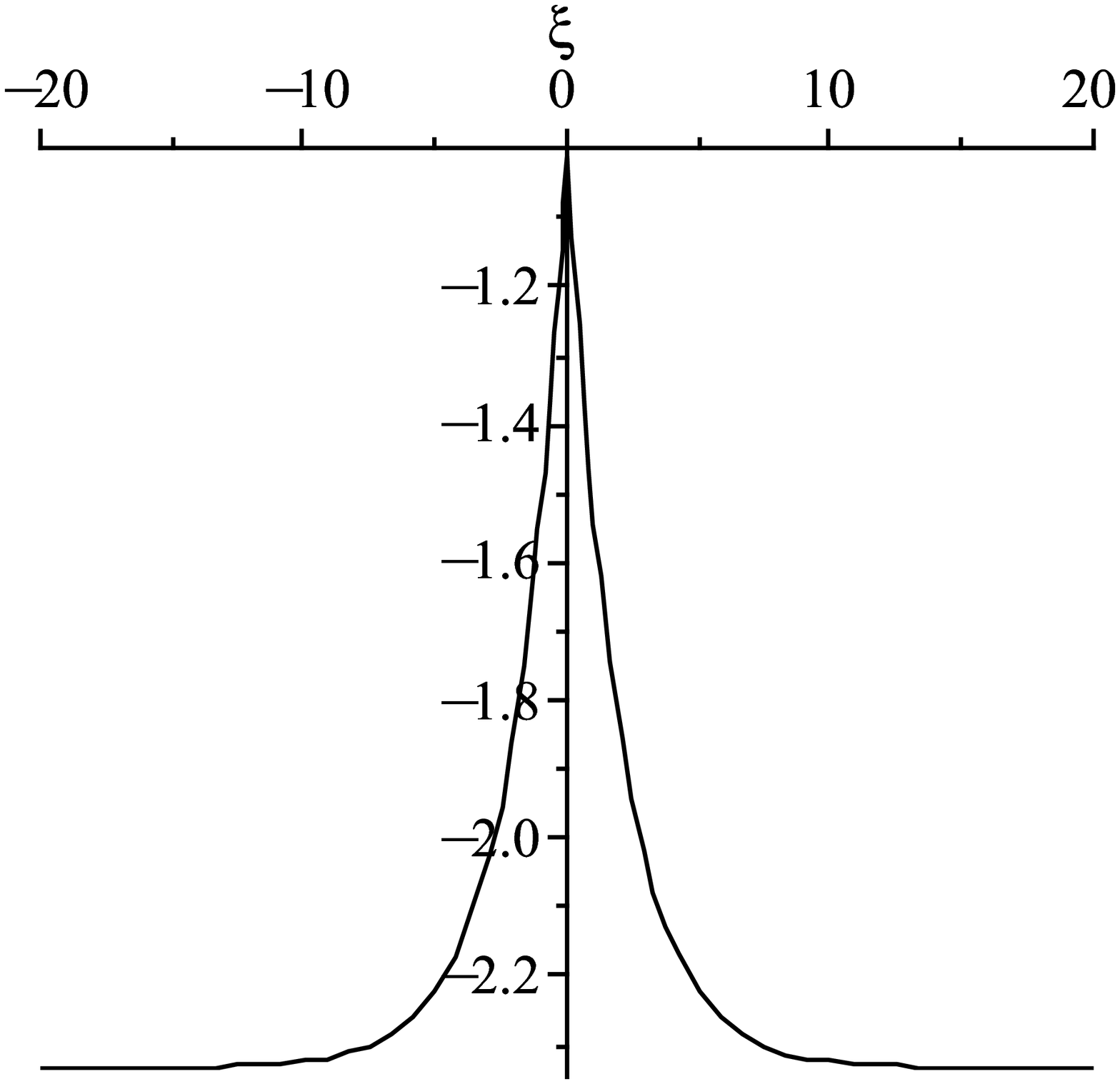}}\hspace{0.2\linewidth}
\subfloat[]{\includegraphics[height=1.5in,width=1.5in]{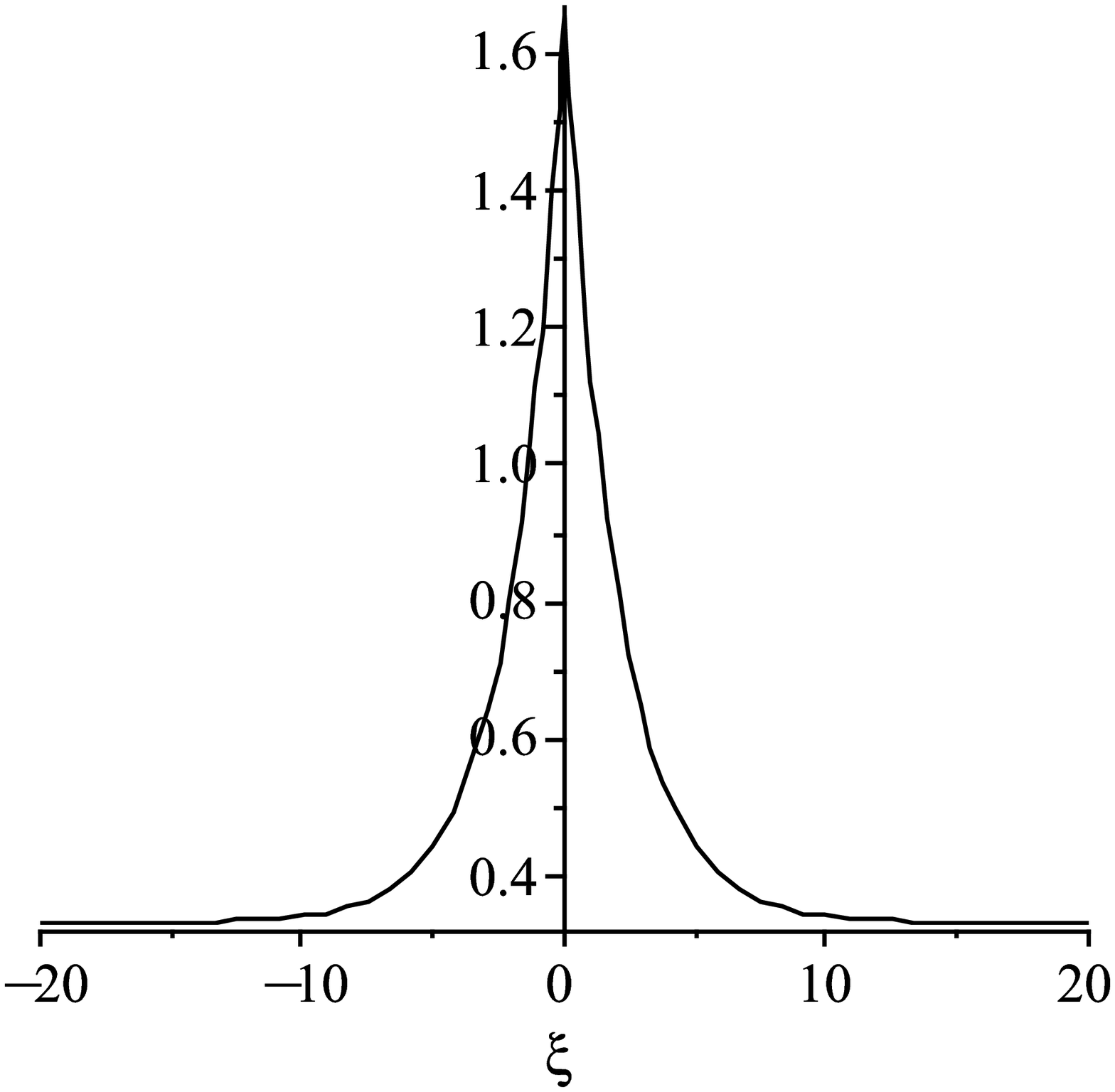}}
\caption{The peakons for Eq.(\ref{eq1.1}). (a) $c=-1$ ; (b)
$c=2$.}\label{f4}
\end{figure}

Note the following facts: when $g_2(c)<g<g_1(c)$ and $g$ tends to
$g_2(c)$, the limit curve of such homoclinic orbit of system
(\ref{eq2.3}) is a triangle with the following three line segments
(see Fig.\ref{f2}(b)):
\begin{equation}
\label{eq3.14} y=\pm \frac{1}{2}(\varphi-c+\frac{4}{3})\quad for
\quad \varphi_{1-}\leq \varphi\leq \varphi_2 ,
\end{equation}
 and

\begin{equation}
\label{eq3.15} \varphi=c \quad for \quad  -\frac{2}{3}\leq y\leq
\frac{2}{3}.
\end{equation}
Let us have $g_2(c)<g<g_1(c)$ and $g$ tends to $g_2(c)$; then we
obtain that
\begin{equation}
\label{eq3.16} \varphi(\xi) = \frac{4}{3} \exp( {-\frac{1}{2}
|\xi|})+c-\frac{4}{3},
\end{equation}
which implies that for arbitrary constant $c\neq 0$,
Eq.(\ref{eq1.1}) has peakons
\begin{equation}
\label{eq3.17} u(x,t) = \frac{4}{3} \exp( {-\frac{1}{2}
|x-ct|})+(c-\frac{4}{3}).
\end{equation}
Obviously, $u$ has peaks at $x-ct=0$. We show graphs of the peakons
in Fig.\ref{f4} under some parameter conditions.

\begin{remark}

(1) If we take $c=\frac{4}{3}$ in (\ref{eq3.17}), then we can see
that (\ref{eq3.17}) agrees with the result in  \cite {3}.

(2) In the phase portaits, the triangle curve  corresponds to a
peakon solution.

\end{remark}

 We have the following lemma, similar to Lemma \ref{le3.1}, which indicates the relationship of periodic
 wave solutions for Eq.(\ref{eq1.1})
and periodic orbits of system (\ref{eq2.3}).
\begin{lemma}
\label{le3.2} Assume that $\Gamma$ is a periodic orbit of system
(\ref{eq2.3}) and that its parameter expression is
$\varphi=\varphi(\xi)$ and $y=y(\xi)$; then $u=\varphi(\xi)$ with
$\xi=x-ct$ is a periodic wave solution for Eq.(\ref{eq1.1}).
\end{lemma}

\begin{figure}[h]
\centering
\subfloat[]{\includegraphics[height=1.5in,width=1.5in]{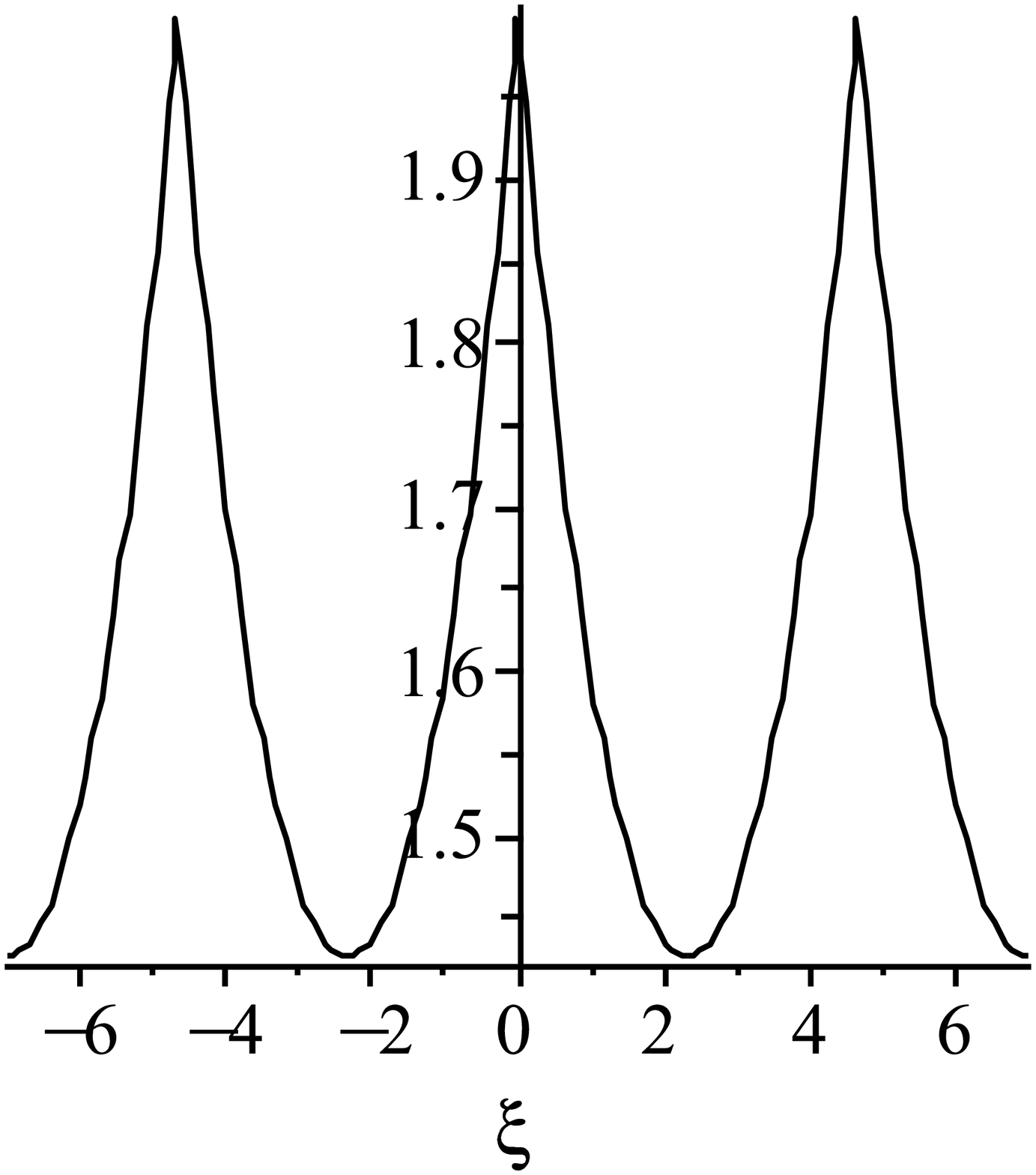}}\hspace{0.2\linewidth}
\subfloat[]{\includegraphics[height=1.5in,width=1.5in]{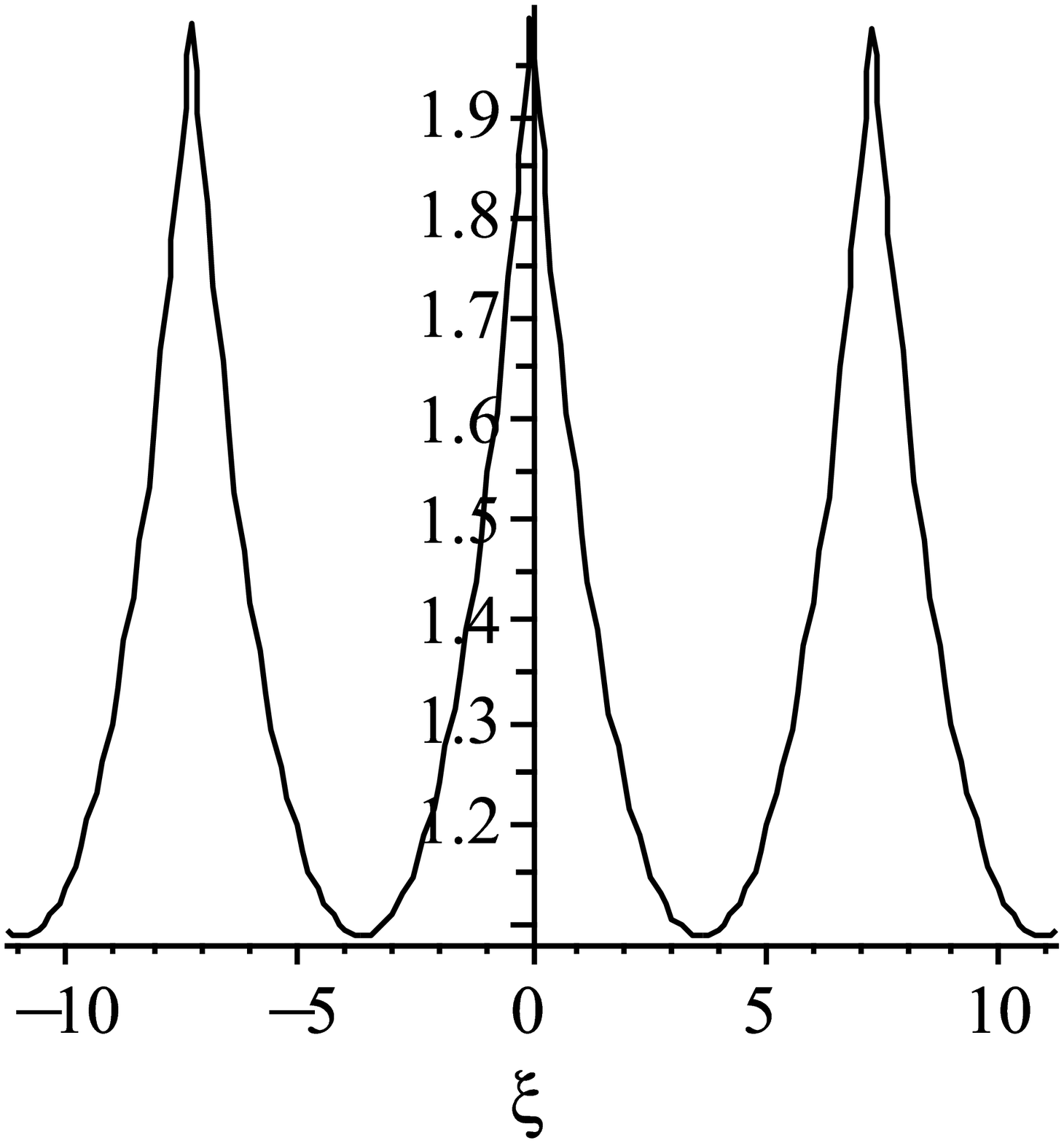}}\\
\subfloat[]{\includegraphics[height=1.5in,width=1.5in]{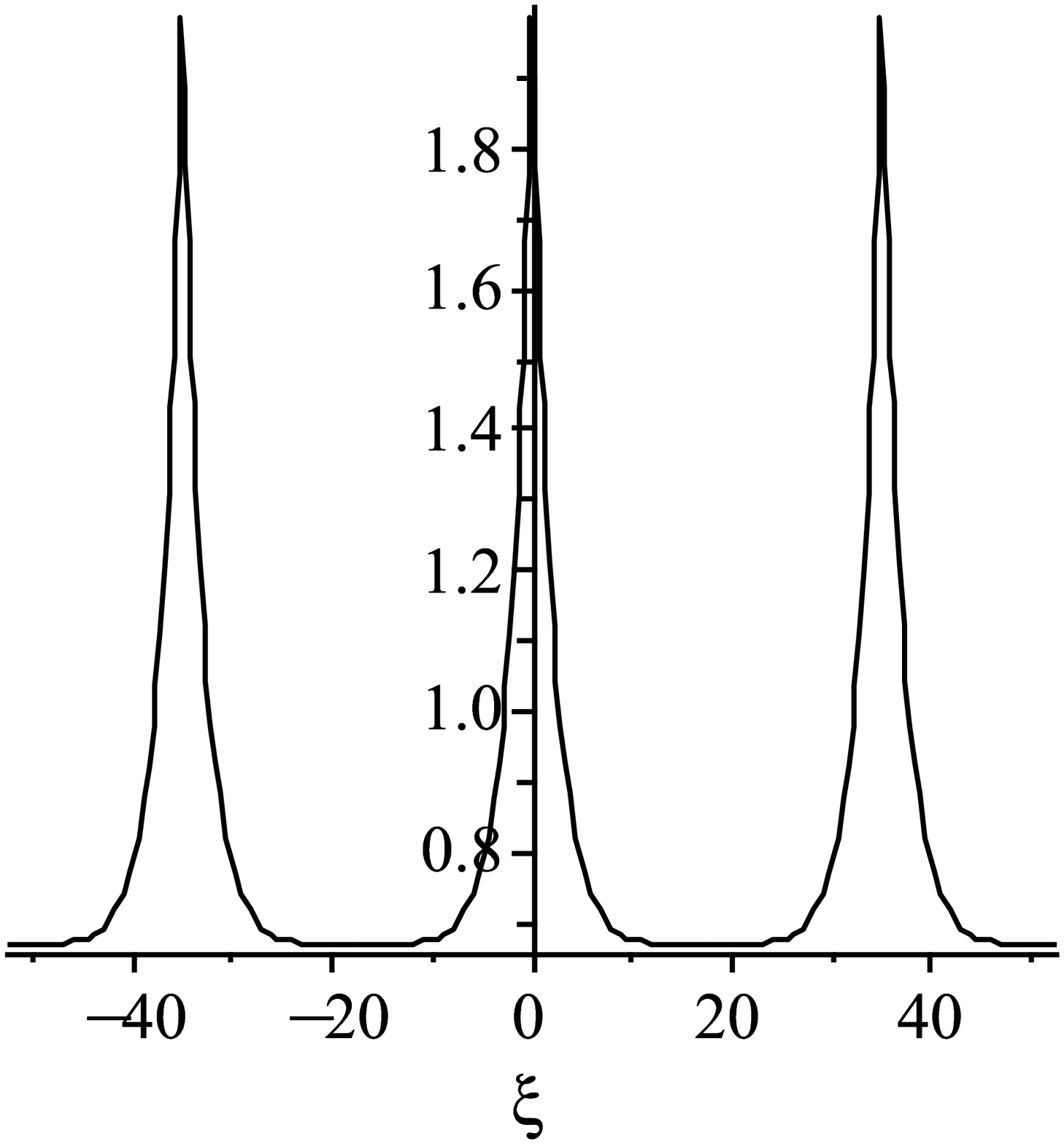}}\hspace{0.2\linewidth}
\subfloat[]{\includegraphics[height=1.5in,width=1.5in]{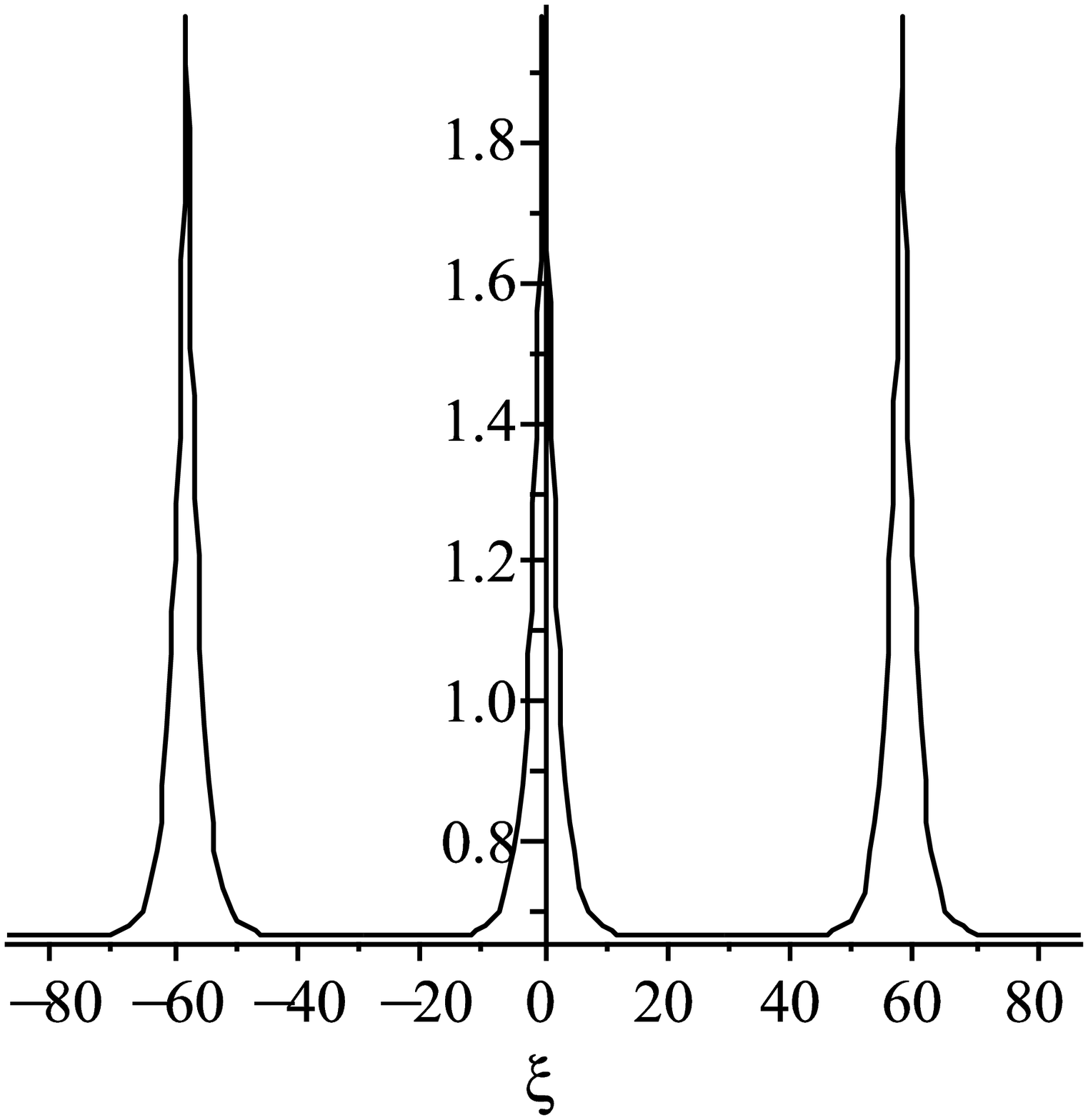}}\\
\caption{The periodic cusp wave solutions for Eq.(\ref{eq1.1}). (a)
$c=2$, $g=0.3$; (b) $c=2$, $g=0.4$; (c) $c=2$, $g=0.4444444$ ; (d)
$c=2$, $g=0.444444444444$.}\label{f5}
\end{figure}

In Fig.\ref{f2}(c), the periodic orbit can be expressed as
\begin{equation}
\label{eq3.18} y=\pm \sqrt{\frac{1}{4}\varphi^2 -
(\frac{1}{2}c-\frac{2}{3})\varphi -\frac{1}{4} c^2+ \frac{1}{3}c+ g}
\quad for \quad \varphi _{2-}\leq \varphi \leq c,
\end{equation}
and
\begin{equation}
\label{eq3.19}\varphi=c \quad for \quad y_{1-}\leq y \leq y_{1+},
\end{equation}
where
\begin{equation}
\label{eq3.20 } \varphi
_{2-}=\frac{1}{3}(-4+3c+\sqrt{2(9c^2-18c+8-18g)} ).
\end{equation}
Substituting (\ref{eq3.18}) into the first equation of system
(\ref{eq2.3}) and integrating along the periodic orbit, we have

\begin{equation}
\label{eq3.21} {\int_\varphi^c {\frac{1}{\sqrt{\varphi^2 -
(2c-\frac{8}{3})\varphi - c^2+ \frac{4}{3}c+4 g}} }ds =
-\frac{1}{2}\xi} \quad for \quad \xi<0,
\end{equation}
and
\begin{equation}
\label{eq3.22} {\int_\varphi^c {\frac{1}{\sqrt{\varphi^2 -
(2c-\frac{8}{3})\varphi - c^2+ \frac{4}{3}c+4 g}} }ds =
\frac{1}{2}\xi} \quad for \quad \xi>0.
\end{equation}
It follows from (\ref{eq3.21}) and (\ref{eq3.22}) that
\begin{equation}
\label{eq3.23}
 \varphi
 (\xi)=l_+
 \exp({-\frac{1}{2}|\xi|})+l_-\exp({\frac{1}{2}|\xi|})+(c-\frac{4}{3})
 \quad for \quad \varphi _{2-}\leq \varphi \leq c,
\end{equation}
where
\begin{equation}
\label{eq3.24} l_\pm=\frac{1}{6}(4\pm 3\sqrt{4g+4c-2c^2}).
\end{equation}
Let
\begin{equation}
\label{eq3.25} T=2|\ln(\varphi _{2-}-c+\frac{4}{3})-\ln(2l_{-})|.
\end{equation}
Then
\begin{equation}
\label{eq3.26} u(x,t)=\varphi(x-ct-2nT) \quad for \quad
(2n-1)T<x-ct<(2n+1)T,
\end{equation}
are periodic cusp wave solutions for Eq.(\ref{eq1.1}) with $2T$
period. Clearly, when $g_3(c)<g<g_2(c)$ and $g\rightarrow g_2(c)$,
$T\rightarrow \infty$, $l_+\rightarrow \frac{3}{4}$, $l_-\rightarrow
0$, and $u(x,t)$ in (\ref{eq3.26}) tends to

\begin{equation}
\label{eq3.27}u(x,t) = \frac{4}{3} \exp( {-\frac{1}{2}
|x-ct|})+(c-\frac{4}{3}).
\end{equation}
(\ref{eq3.27}) is identical with (\ref{eq3.17}). The graphs of some
periodic waves for Eq.(\ref{eq1.1}) are shown in Fig.\ref{f5} under
some parameter conditions. From Fig.\ref{f5} we can see that
$g_3(c)<g<g_2(c)$ and $g$ tends to $g_2(c)$, the periodic cusp wave
solutions tend to peakons.

\begin{remark}
(1) In the phase portaits, the semi-elliptic closed curve with one
side on the singular line $\varphi=c$ corresponds to a periodic cusp
wave solution.

(2) In \cite{4}, we dealt with the case $g\leq g_1(c)$, and obtained
the kink-like and antikink-like wave solutions for Eq.(\ref{eq1.1}).
\end{remark}

\section{Conclusion}

In this work,  by using the bifurcation method, we obtain the
analytic expressions for solitons, peakons and periodic wave
solutions for the Fornberg-Whitham equation, given as (\ref{eq3.6}),
(\ref{eq3.17}) and (\ref{eq3.26}), respectively. We also show the
relationships among the solitons, peakons and periodic cusp wave
solutions.

\end{document}